\documentclass[traditabstract]{aa}
\usepackage{graphicx}
\usepackage{txfonts}
\usepackage{color}
\usepackage{natbib}
\usepackage{epsfig}
\usepackage{hyperref}
\usepackage{psfrag}
\usepackage{lineno}
\begin{document}
   \title{Constraining the location of rapid gamma-ray flares in the FSRQ 3C 273}

   \titlerunning{GeV Variability in 3C 273 }

   \author{
          B. Rani \inst{1}\thanks{Member of the International Max Planck Research School (IMPRS) for Astronomy and
Astrophysics at the Universities of Bonn and Cologne}
          \and B. Lott \inst{2}
          \and T. P.\ Krichbaum \inst{1}
          \and L. Fuhrmann \inst{1}
          \and J.A. Zensus \inst{1} \\
         }
   \institute{
              Max-Planck-Institut f$\ddot{u}$r Radioastronomie (MPIfR), Auf dem H{\"u}gel 69, D-53121 Bonn, Germany
           \and
              Universit{\'e} Bordeaux 1, CNRS/IN2p3, Centre d'Etudes Nucl{\'e}aires de Bordeaux Gradignan,
            33175 Gradignan, France
          }

   \date{Received ---------; accepted ----------}

 \abstract 
{
We present a $\gamma$-ray photon flux and spectral variability study of the flat-spectrum radio quasar 3C 273 
over a rapid flaring activity period between September 2009 to April 2010.  
Five major flares are observed in the source during this period. The most rapid flare observed in the source has 
a flux doubling time of 1.1 hr.
The rapid $\gamma$-ray flares allow us to constrain the location and size of the $\gamma$-ray emission region in 
the source. The $\gamma \gamma$-opacity constrains the Doppler factor, $\delta_{\gamma} \geq$ 10 for the 
highest energy (15 GeV) photon observed by the {\it Fermi}-Large Area Telescope (LAT). 
Causality arguments constrain the size of the emission region to 1.6$\times 10^{15}$ cm. 
The $\gamma$-ray spectra measured over this period show clear deviations from a simple power law with a 
break in 1-2 GeV energy range. We discuss possible explanations for the origin of the $\gamma$-ray spectral breaks. 
Our study suggests that the $\gamma$-ray emission region in 3C 273 is located within the broad line region ($<$1.6 pc). 
The spectral behavior and temporal characteristics of the individual flares indicate the presence of multiple 
shock scenarios  at the base of the jet.
}


\maketitle

\section{Introduction}
3C 273 is classified as a flat spectrum radio quasar (FSRQ) at a redshift, z = 0.158 \citep{strauss1992}. 
It was the first discovered quasar \citep{schmidt1963}, and the first extragalactic source detected in 
$\gamma$-ray band, by {\it COS-B} \citep{swanenburg1978}. The source is categorized as a low synchrotron peaked blazar 
(LSP) by \citet{abdo2010a}. The broadband spectrum, the correlation among multi-frequency flares and 
the VLBI jet kinematics of the source were extensively studied in the past \citep[e.g.][ and references therein]
{krichbaum2001, rani2010, rani2011, soldi2008, savolainen2010, abdo_3c273}.

After the first detection in $\gamma$-rays by {\it COS-B}, 3C 273 was also observed by 
EGRET in 1999 \citep[3EG J1229+0210 in][]{hartman1999} with an average flux of 
F(E$>$100 MeV) = 0.18$\times$10$^{-6}$ ph cm$^{-2}$ s$^{-1}$. 
\citet{krichbaum2001} noticed a correlation between jet component ejection and EGRET $\gamma$-ray 
flux variations, which constrained the location of the $\gamma$-ray emission region to $\sim$2000 Schwarzschild radii 
($< 0.2$ pc) from the central black hole (BH). 
{\it Fermi}--Large Area Telescope (LAT) detected the source in the $\gamma$-ray band from the 
beginning of observations in 2008 \citep{abdo2010a}. The source was observed in a bright outburst phase in the 
beginning of July 2009. The first flaring event lasted for $\sim$10 days in August 2009 \citep{bastieri2009}. 
The source exhibited another $\gamma$-ray outburst in September 2009 with two bright flares from September 15 to 19, 2009 
and from September 20 to 23, 2009.   \citet{abdo_3c273} discussed the flare characteristics and spectral behavior 
during this outburst period. In fact, before the brightest $\gamma$-ray flare observed in 3C 454.3 in December 2009 
\citep{escande2009,abdo_3c454}, 3C 273 was the brightest extragalactic non $\gamma$-ray burst source observed by 
{\it Fermi}-LAT.
Although the $\gamma$-ray spectrum of 3C 273 integrated over the first 6 months of LAT observations showed a significant break around 
1.5 GeV \citep{abdo_spectra}, the spectra integrated over shorter periods of time in September 2009 did not provide 
indication for such curvature \citep{abdo_3c273}.

The second flare during September 20 to 30, 2009  displayed a first short peak and was followed by a period of 
high activity indicating either 
a series of substructures or a long decay tail \citep{abdo_3c273}. The source continued the rapid flaring activity  
until April 2010 and later was observed in a quiescent state, which is still continuing.  
We refer to the flaring interval between August 2009 to April 2010 as the 
high activity period. 
Thanks to the high flux during August 2009 to April 2010, we are able to obtain light curves with good  
statistics  using a time binning of only 3 hr, allowing us to investigate the evolution of these outbursts.
The focus of this study is to probe the location and size of the $\gamma$-ray emission region in 3C 273 by investigating its 
intra-day variability and associated spectral changes in the $\gamma$-ray band.

This paper is structured as follows. Section 2 provides a brief
description of observations and data reduction. In Section 3, we report our
results. We discuss our results in Section 4. Conclusions are given in Section 5.

\section{Observations and data reduction}
\label{obs}
The $\gamma$-ray flux and spectral variations of the source are investigated using the 
{\it Fermi}-LAT data. The $\gamma$-ray data employed here were 
collected during a time period between JD = 2454683 (August 04, 2008) to 2455889 (November 23, 2011) in survey mode by  
{\it Fermi}-LAT. We analyze the LAT data using the standard 
ScienceTools (software version v9.23.1) and instrument response functions 
P7V6\footnote{http://fermi.gsfc.nasa.gov/ssc/data/analysis/scitools/overview.html}. 
Photons in the Source event class are selected for this analysis.  
We select $\gamma$-ray photons with zenith angles less than $100^{\circ}$ to greatly reduce contamination from $\gamma$-rays
produced by cosmic-ray interactions in the upper atmosphere.
The diffuse emission from our Galaxy 
is modeled using a spatial model (gal$\_$2yearp7v6$\_$v0.fits) which is refined by the {\it Fermi}-LAT 
data taken during the first two years of operation. The extragalactic diffuse and residual 
instrumental backgrounds are modeled as an isotropic component (isotropic$\_$p7v6source.txt),
and are provided with the data analysis tools. 
We analyze a region of interest of 10$^\circ$ in radius, centered 
at the position of the $\gamma$-ray source associated with 3C 273 using a 
maximum-likelihood algorithm\footnote{http://fermi.gsfc.nasa.gov/ssc/data/analysis/scitools/likelihood$\_$tutorial.html} 
\citep{mattox1996}.
In the model for the $\gamma$-ray emission from the region, we include all 
the 30 sources within 10$^\circ$ with their model parameters fixed to their catalog values except 
for 3C 279 and  MG1 J123931+0443. The other sources are reported as being not significantly variable in the 2FGL 
catalog \citep[see][]{2fgl_cat}. For 3C 279 and  MG1 J123931+0443, we set all 
model parameters free.

We investigate the source variability by producing light curves by likelihood analysis with different  
time binnings (3 hr, 6 hr, 12 hr, 1 day and 1 week) and over different energy ranges 
(E $>$ 100 MeV, E $<$ 1 GeV and E $>$ 1 GeV). The light curves are produced by 
modeling the spectra for each time bin and energy range by a simple power law  which is sufficient for the 
relatively short time ranges considered here. 
Also, the statistical uncertainties on the power law (PL) indices are smaller than those 
obtained from broken-power law (BPL) fits. 

We also compute photon flux light curves above the ``de-correlation energy", E$_0$ \citep{lott2012}, which  minimizes the
correlations between integrated photon flux and photon index ($\Gamma$).
Over the course of $\sim$3 years of observations, we find $E_0$ = 167 MeV.  
We generated the constant uncertainty (15$\%$) 
light curve above $E_0$ through the adaptive binning method following \citet{lott2012}. 
The estimated systematic uncertainty on the flux is 10$\%$ at 100 MeV, 5$\%$ at 500 MeV, and 20$\%$ at 10 
GeV \citep{ackermann2012}.

We perform the spectral analysis by fitting the $\gamma$-ray spectra with multiple models over the whole energy 
range covered by {\it Fermi}-LAT above 100 MeV. The different spectral forms are  simple power law  
[PL, $N(E) = N_{0} (E)^{-\Gamma}$, $N_0$ : Prefactor and $\Gamma$ : power-law index], 
 smoothed-broken power law 
[SBPL, $N(E) = N_0 E^{-\Gamma_1} \big (1 + (E/E_{break})^{-(\Gamma_1 - \Gamma_2)/b} \big )^{-b}$, 
$\Gamma_1$, $\Gamma_2$ : the two power-law indices and $E_{Break}$ : break 
energy, $b$ : smoothing parameters set to 0.1] 
and a log-parabola function [logP, $N(E) = N_0 (E/E_{P})^{-(\alpha + \beta ~log(E/E_{P})}$, $\alpha$ : power-law index, 
$\beta$ : index parameter that characterizes the spectral curvature and 
$E_P$ : the peak energy fixed at 1 GeV].  
We also examine the spectral behavior over the whole energy range with a simple PL 
model that we fit over equally spaced logarithmic energy bins, with $\Gamma$ kept constant and equal to the value 
obtained by fitting over the whole energy range.

\begin{figure*}
\center
\epsfig{figure= 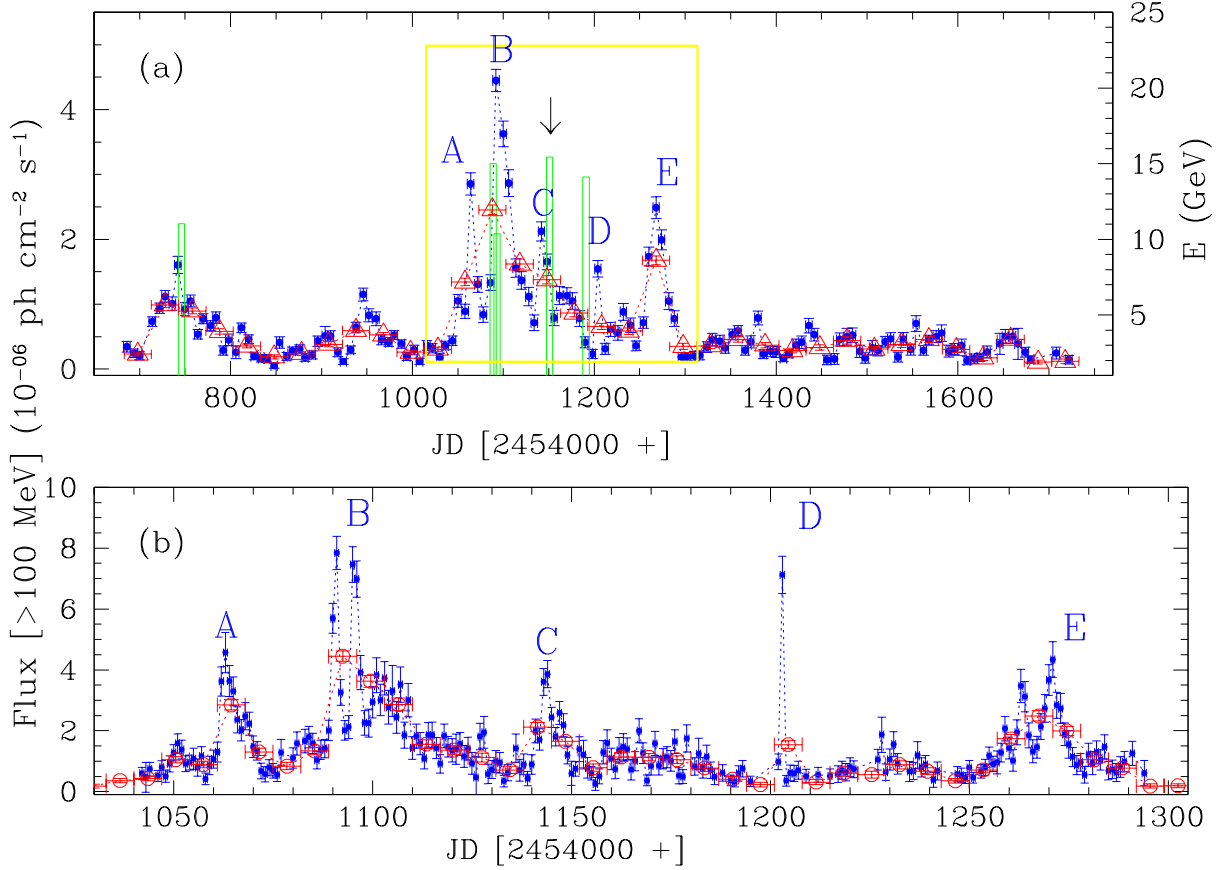,scale=0.6,angle=-90}
\caption{(a) Weekly (blue circles) and monthly (red open symbols) averaged flux (E$>$100 MeV) light curves 
of 3C 273 measured by {\it Fermi}-LAT since its launch. The green histogram represents the arrival time distribution of 
E $>$ 10 GeV photons associated with 3C 273.  The yellow box represents the high $\gamma$-ray flux activity 
period.} (b) 1-day averaged light curve of the source over the high activity period 
within the  box shown in the top of the figure.      
\label{lc_flx}
\end{figure*}

\section{Analysis and Results}

\subsection{Temporal Characteristics}

\subsubsection{Light curves}
\label{sec_lc}
The source has been exhibiting significant flux variability throughout the whole {\it Fermi} mission \citep{abdo_flare}. 
Fig. \ref{lc_flx} (a) displays the weekly averaged flux $F_{100}$ (F[E $>$ 100 MeV] in units of 10$^{-6}$ 
ph cm$^{-2}$ s$^{-1}$) light curve of the source. 
A sequence of flares is observed in the source between JD$^{\prime}$\footnote{JD$^{\prime}$ = JD -2454000} = 1042 
to 1320 (August 2009 to April 2010) and no substantial flux variability has been observed later. 
The source seems to have been in a quiescent state since April 2010, which is still continuing.  The high activity period 
is visualized by a box in Fig. \ref{lc_flx}. 
Fig. \ref{lc_flx} (b) shows the 1-day averaged $F_{100}$ light curve over this period. Five major flares 
labeled `A' to `E' are observed in the source during this period (see Fig. \ref{lc_flx} b). Interestingly, 
the major flares are separated from each other by roughly 50 days.  
The high flux level during the flaring periods allow us to investigate 
the evolution of outbursts with a temporal resolution as fine as 3 hr (corresponding to one sky scan of the LAT).

Fig. \ref{lc_flx} (a) shows the arrival time distribution of $\gamma$-ray photons with energies greater than 10 GeV. 
During the $\sim$3 years of observations, the highest energy photon associated with 3C 273 was detected at
JD$^{\prime}$ =  1151 with an energy of 15.4 GeV. 
This photon is  converted in the front section of the LAT tracker, where the angular resolution is best \citep{ackermann2012}.
The reconstructed arrival direction of the photon
is 0.07$^{\circ}$ away from 3C 273.  
Based on our model fit of the epoch which contains the highest-energy photon, we find the probability that the
photon was associated with 3C 273 (as opposed to all other sources in the model including the diffuse emission
and nearby point sources) is 99.98$\%$, which is $>$3$\sigma$. Interestingly, the highest  energy photon 
is observed during the flaring period.

Flare A had a duration of $\sim$12 days with a peak flux of $\sim5.1 \times 10^{-6}$ 
ph cm$^{-2}$ s$^{-1}$. Flare B consisted of two main sub-flares, each one having a total duration of 
$\sim$2 days and a peak flux of $\sim12 \times 10^{-6}$ ph cm$^{-2}$ s$^{-1}$. The
first sub-flare showed a nice smooth time profile, suggesting that it is well resolved; the second sub-flare, 
after a first short peak, shows a series of high spikes that indicate substructures in  
a long decay tail. A detailed study of flux and spectral variations of these two outbursts is given 
in \citet{abdo_3c273}. 
Flare C is recorded $\sim$50 days later than flare B, and is followed by an extremely rapid flare D and a 
number of sub-flares as flare E.

\begin{figure}
\epsfig{figure= 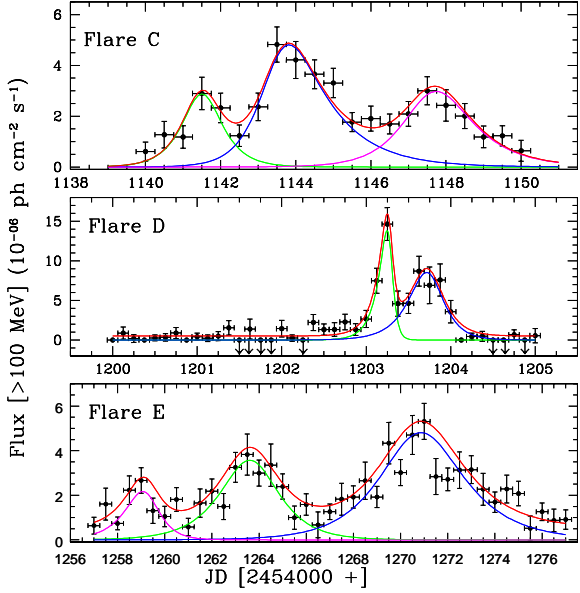,scale=0.4,angle=0}
\caption{Light curves of 3C 273 above 100 MeV with time binning of 12 hr (flare C and E) and 3 hr (flare D). 
The lines are the fitted sub-flaring components (see Table \ref{tab1}).  }
\label{flare_model}
\end{figure}

We further investigate the evolution of these flares with finer time resolution  
depending upon the $\gamma$-ray brightness of the source. The light curves for $F_{100}$,  with 3 hr and 12 hr 
time binning for the individual flares are shown in 
Fig. \ref{flare_model}. A fit consisting of a slowly varying background and rapid sub-flaring components is performed 
for each individual flare.  The slowly varying background is roughly approximated by a flux 
value = 0.42 $\times$ 10$^{-6}$ ph cm$^{-2}$ s$^{-1}$. 
 Each component is fitted by a function of the form : 
\begin{equation}
F(t) = 2~F_0 \big [ e^{(t_0 - t)/T_r} + e^{(t - t_0)/T_f} \big ]^{-1}
\end{equation}
where T$_r$ and T$_f$ are the rise and decay times, respectively, and F$_0$ is the flux at t$_0$ representing
approximately the flare amplitude. The solid curves in Fig. \ref{flare_model} represent the fitted flare components 
and the fitted parameters for each sub-component are given in Table \ref{tab1}. The doubling times of 
the individual flares is given in the last column of Table \ref{tab1}. Note that the flares have similar 
halving times because of comparable rise and decay timescales.

\begin{figure}
\epsfig{figure= 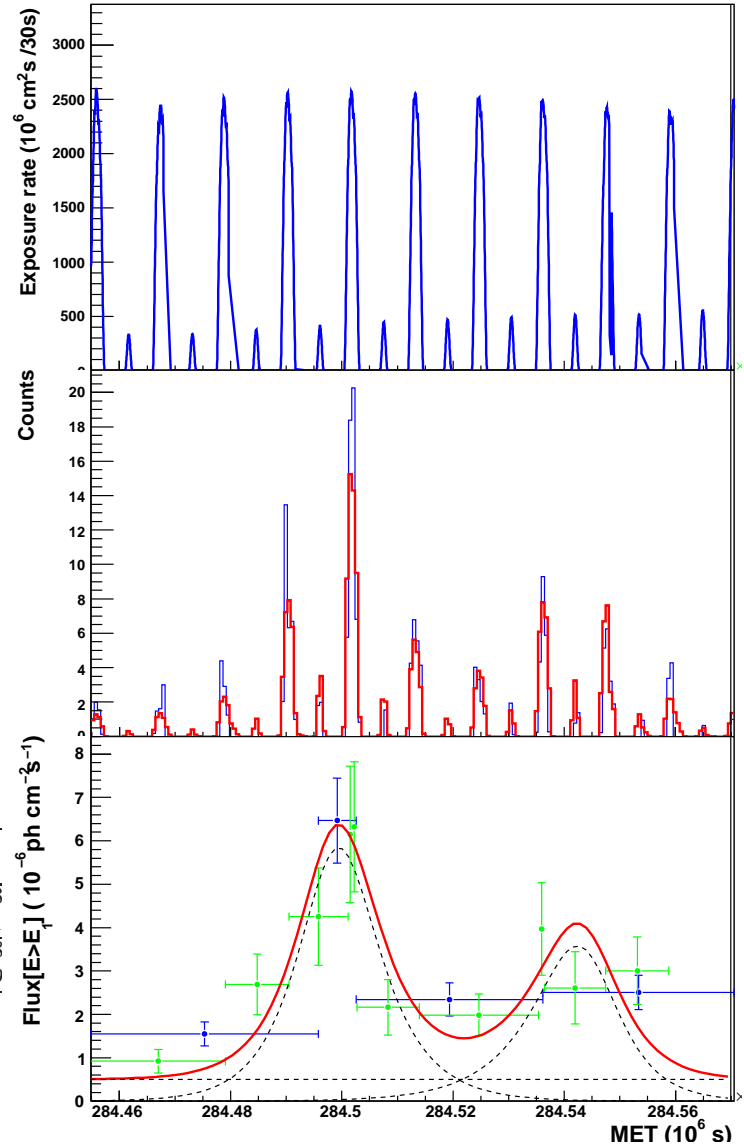,scale=0.6,angle=0}
\caption{Top: Exposure rate during flare D. Middle: Predicted (red) and observed (blue) 
source photon detection rates. Bottom:  Photon flux light curve of flare D. The red curve shows the results of the 
time-domain-unbinned analysis, and the blue (15$\%$) and green (25$\%$) symbols represent the adaptive-binning analysis 
results.  }
\label{flare_D}
\end{figure}

The short duration of flare D, with 
timescales similar to the orbital period ($\sim$96 min) of {\it Fermi}, makes the standard analysis, where the photons are binned in time,  
inappropriate for a detailed temporal characterization of the flare.  A different, unbinned (in the time domain) 
method taking into account the strongly varying rate of accumulation of exposure associated with the survey mode has been employed 
instead for this purpose. In this method, based on the same log-likelihood approach as described in \citet{lott2012},  a  
time-dependent function $f(t,E) = F(t) \times N(E)$  where $F(t)$ is given in Eq. 1 and $N(E)$ is a power-law 
distribution (neglecting any spectral changes during  the flare), has been fitted to the {\sl unbinned data}.
The results are plotted in Fig. \ref{flare_D}, showing from top to bottom the exposure rate, the comparison between 
the predicted source-photon detection rate and that estimated from the data, and the flux light curve. The 
parameters for $F(t)$ are given in Table \ref{tab1}.

During the first component of the flare D  
the source brightens to $F_{100}$ $\sim$12  with a flux doubling time of 1.1 hr. 
The second component of flare D 
has a peak flux value $F_{100}$ $\sim$7  with a flux doubling time of 1.5 hr. 
During flare D, the Moon was close to the region of interest of 3C 273. 
We therefore checked for the possible contribution from the Moon. We found that the Moon was closest to source on 
January 07, 2010 with an angular distance of 11$^{\circ}$, which is larger than the region of interest of the source,  
and therefore unlikely to cause any significant contamination in the observed rapid flux variations.

\begin{table*}
\center
\caption{Fitted parameters of flares}
\begin{tabular}{ccccccc} \hline
Main- & Sub-     & T$_r$         &  T$_f$        & t$_0$         & F$_0$               & Doubling  \\
Flare & Flare    & (day)         &  (day)        & JD$^{\prime}$           & 10$^{-6}$ ph cm$^{-2}$ s$^{-1}$ & Time (hr)$^{a}$    \\\hline   
C     & 1        & 0.34$\pm$0.12 &0.39$\pm$0.13  &1141.52$\pm$0.11 &2.31$\pm$0.13        & 5.60     \\  
      & 2        & 0.56$\pm$0.13 &1.01$\pm$0.10  &1143.71$\pm$0.23 &4.11$\pm$0.12        & 9.30      \\
      & 3        & 0.68$\pm$0.12 &0.87$\pm$0.18  &1147.61$\pm$0.64 &2.20$\pm$0.13        & 11.31     \\\hline
D$^{b}$& 1       & 0.07$\pm$0.02 &0.08$\pm$0.02  &1203.31$\pm$0.04 &12.30$\pm$2.83       & 1.16     \\
      & 2        & 0.09$\pm$0.04 &0.07$\pm$0.02  &1203.83$\pm$0.05 &7.41$\pm$3.09        & 1.54      \\\hline
E     & 1        & 0.79$\pm$0.13 &0.53$\pm$0.11  &1259.22$\pm$0.24 &2.13$\pm$0.10        & 13.10     \\
      & 2        & 1.16$\pm$0.16 &1.10$\pm$0.21  &1263.63$\pm$0.35 &3.43$\pm$0.09        & 19.32     \\
      & 3        & 1.47$\pm$0.13 &1.68$\pm$0.18  &1270.72$\pm$0.31 &4.14$\pm$0.12        & 24.43     \\\hline
\end{tabular} \\
$^{a}$ : doubling time = $T_r \times ln2$ \\
$^{b}$ : In estimating the photon detection rate from the data, the relative fluxes of the source and background 
have been taken from the binned-analysis results. This estimated rate is not strongly dependent on these relative 
fluxes, though, due to the different spatial distributions of the source and background components. 
\label{tab1}
\end{table*}

Flare A shows essentially equal rise ($\sim$1.81 day) and decay timescales ($\sim$1.79 day), which implies that 
flare A has a symmetric temporal profile. In contrast to flare A, the components of flare B are 
characterized by decay times ($\sim$2.50 day) longer than the rise time ($\sim$0.50 day)  \citep{abdo_3c273}. 
For flare C, the components have a symmetric 
profile except the middle sub-component (see Table \ref{tab1}). Both the components of flare D have relatively symmetric 
evolution. All the components of flare E are characterized by symmetric temporal profiles. 
Therefore, most of the flares have symmetric profiles except flare B.

\subsubsection{Flux variations at different Energy bands}
\begin{figure*}
\includegraphics[scale=0.6,angle=-90, trim = 120 0 0 0, clip]{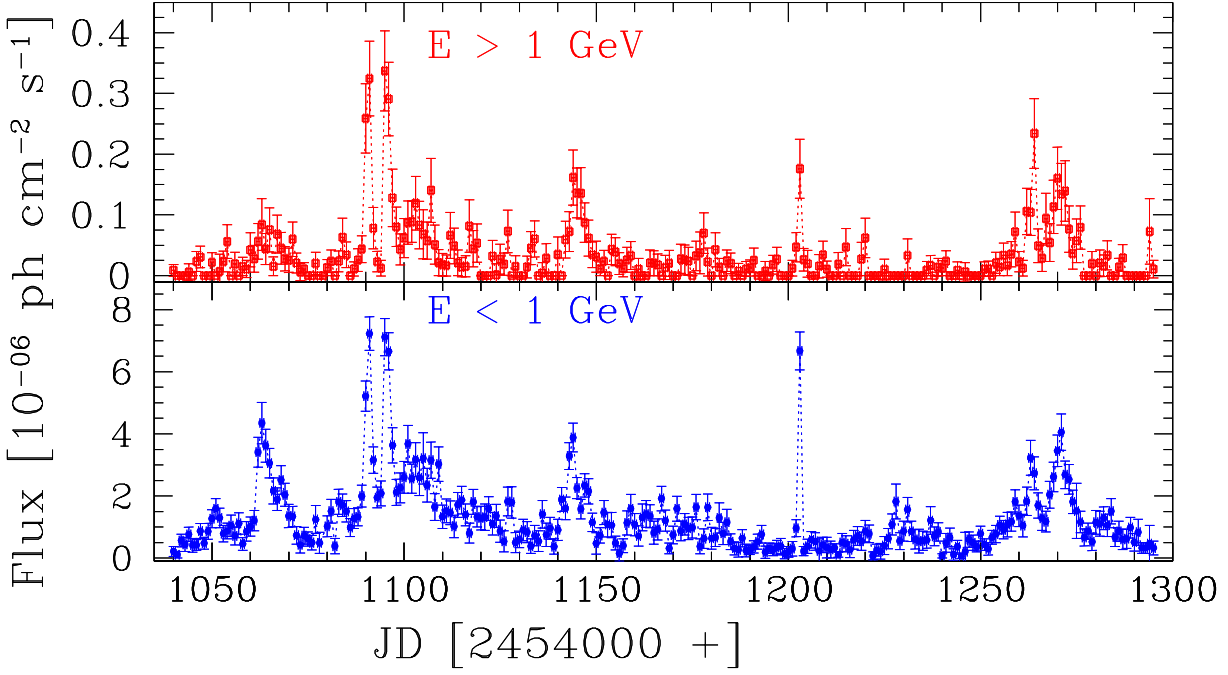}
\caption{ 1-day averaged $\gamma$-ray light curve of the source below (bottom) and above (top) 
1 GeV during the high activity period.  
   }
\label{plot_flx_allE}
\end{figure*}

The flux variations below and above 1 GeV energy during the high activity period are shown in Fig. \ref{plot_flx_allE}. 
Apparently the flux variations below 1 GeV seem to be more pronounced, but we note that the characteristic 
decay timescales of the individual flares are similar 
for the $F_{E<1 GeV}$ and $F_{E>1 GeV}$ light curves using 12 hr time binning. Even with a finer temporal 
resolution of 6 hr for the brightest flare B (component 1 and 2), we find that the decay timescales for 
the $F_{E<1 GeV}$ and $F_{E>1 GeV}$ light curves are not statistically different. The rise and decay timescales, 
the peak flux value and the respective time of peak of the flare for the 6 hr binned light curves of these 
flares at different energy scales are given in Table \ref{tab_Ediff}.

\begin{figure}
\includegraphics[scale=0.3,angle=0]{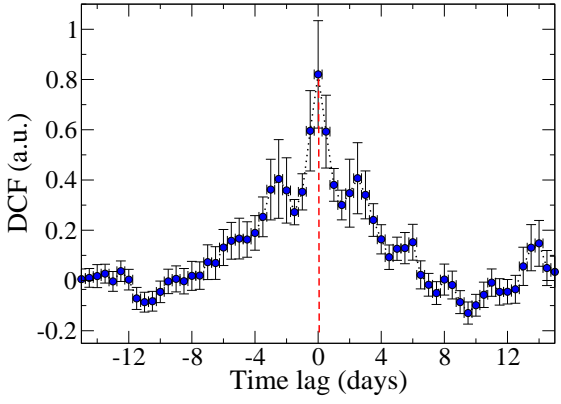}
\caption{ DCF curve of the 12 hr binned $\gamma$-ray light curves below and above 1 GeV. The 
dotted line marks a possible time lag of 0$\pm$0.25 day. 
   }
\label{plot_dcf_allE}
\end{figure}

To search for a possible time lag and to quantify the correlation among the $\gamma$-ray flux variations below and 
above 1 GeV, we compute the discrete cross-correlation function (DCF) of the two light curves (E below and above 1 GeV) following the method 
described by \citet{edelson1988}. The 12 hr binned DCF analysis curve (Fig. \ref{plot_dcf_allE}) shows a peak 
(DCF = 0.82$\pm$0.21)  at a time lag = 0.00$\pm$0.25 days. This confirms the existence of a significant correlation 
between the flux variations below and above 1 GeV with zero time delay. Due to limited statistics for the more finely binned 
light curves, we can not claim any time delay shorter than our binning interval of 12 hr.

\begin{table}
\center
\caption{Fitted parameters of the bright flares at different energy scales}
\begin{tabular}{ccccccc} \hline
Flare & Sub-flare  & Energy   & T$_r$         &  T$_f$        & t$_0$         & F$_0$                \\
      &            & (GeV)      & (day)         &  (day)        & JD$^{\prime}$           & 10$^{-6}$ ph cm$^{-2}$ s$^{-1}$  \\\hline   
B     &1           & 0.1-1    & 0.76$\pm$0.09 &1.22$\pm$0.14  &1090.92$\pm$0.33 &7.15$\pm$0.90      \\  
      &            & 1-100    & 0.53$\pm$0.16 &1.39$\pm$0.18  &1090.81$\pm$0.21 &0.35$\pm$0.12       \\\hline 
B     &2           & 0.1-1    & 0.68$\pm$0.14 &1.53$\pm$0.21  &1095.02$\pm$0.18 &7.21$\pm$0.78      \\  
      &            & 1-100    & 0.46$\pm$0.17 &1.42$\pm$0.33  &1094.93$\pm$0.19 &0.30$\pm$0.05       \\\hline
\end{tabular} \\
\label{tab_Ediff}
\end{table}

\subsubsection{Photon index variations}
\begin{figure*}
\center
\includegraphics[scale=0.56,angle=-90, trim = 210 0 -10 0, clip]{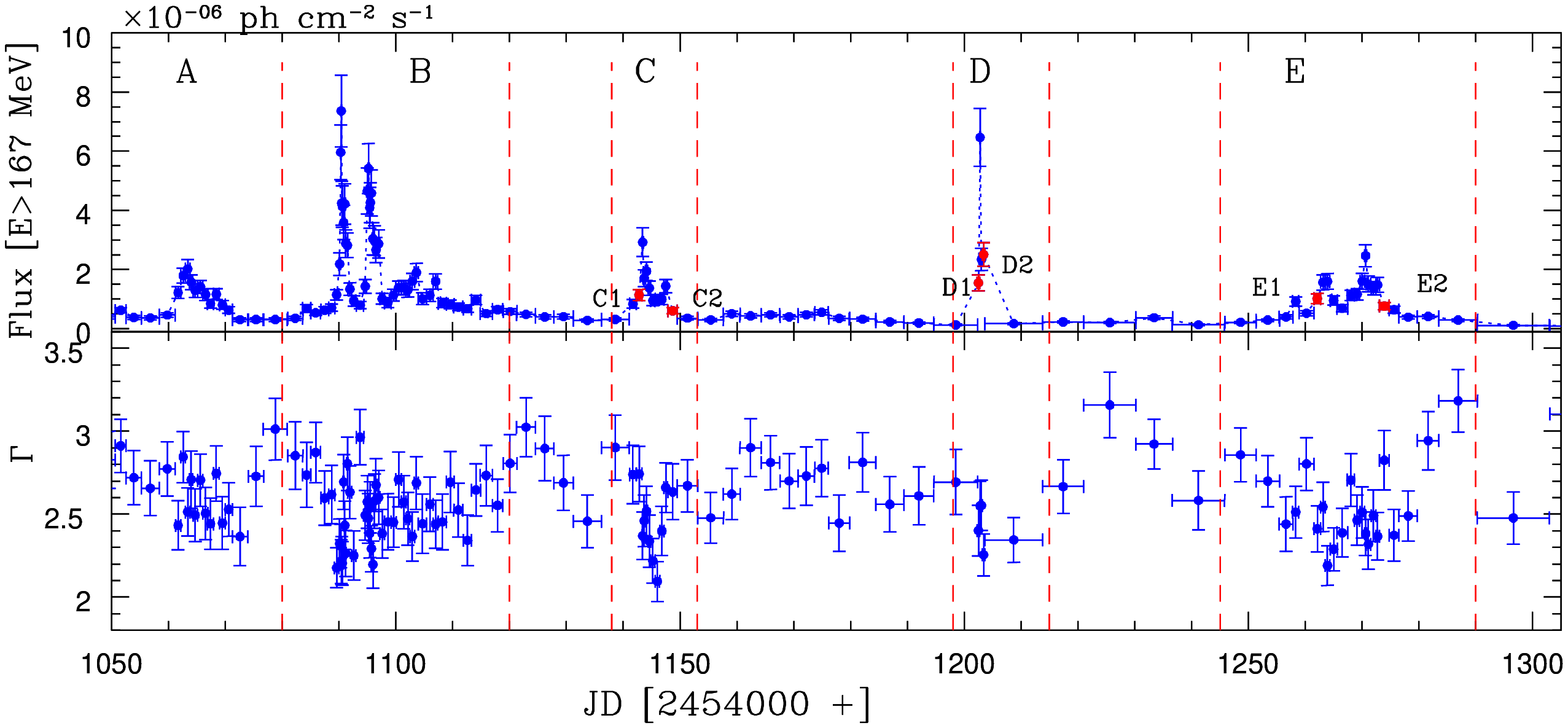}
\includegraphics[scale=0.184]{adp_flareC.eps}
\includegraphics[scale=0.184]{adp_flareD.eps}
\includegraphics[scale=0.184]{adp_flareE.eps}
\caption{ Top panel : The constant uncertainty (15$\%$) light curve of 3C 273 above $E_0$ obtained with the adaptive 
binning method  over the high activity period (see text for details). The lower panel shows the 
variations in $\Gamma$ during this period. The vertical columns separate the different flaring periods. 
Bottom : $\Gamma$ vs. flux above $E_0$ for flare C, D and E. 
   }
\label{flx_pvoit}
\end{figure*}

We investigate the flux variations above the de-correlation energy ($E_0$), which minimizes the correlations 
between integrated photon flux and photon index.  We generate the constant uncertainty (15$\%$) 
light curve above $E_0$ using the adaptive binning method \citep{lott2012}.
Fig. \ref{flx_pvoit} shows the photon flux and index variations during the high activity period between 
JD$^{\prime}$ = 1050 to 1320 at E$>$E$_{0}$.  The different flaring activity periods of the source are separated by vertical columns. 
We note substantial variations in photon spectral index during different modes of flux activity. As we see in 
Fig. \ref{flx_pvoit}, for each flare $\Gamma$ drops from a value of $\sim$(2.9 -- 2.7) (at the beginning of the flare) 
to $\sim$(2.3 -- 2.1) (close to the peak of the flare) and then at the end of the flare the spectrum softens to a photon index 
value $\sim$(3.0 -- 2.8). Therefore, a strong evolution of $\Gamma$ during the different activity states is observed 
in the source.

An alternative approach to investigate the spectral evolution during different $\gamma$-ray flares is to plot variations in 
$\Gamma$  as a function of flux variations.   A counter-clockwise loop is observed during flare A while 
during flare B, the spectral index and flux changes follow a clockwise path \citep{abdo_3c273}.  
In Fig. \ref{flx_pvoit} (bottom), we also plot the photon indices as a function of the $\gamma$-ray fluxes during flares C, D 
and E. As shown, flares C and E suggest a clockwise loop, while flare D follows a counter-clockwise loop.  
A clockwise loop indicates that the flux started to increase at low energy and then propagates to
high energy; the reverse would produce a counter-clockwise path. This may reflect the alternating dominance 
of acceleration and cooling processes.  We note that the subsequent flares loop in opposite sense, starting 
with a counter-clockwise loop for flare A.

\subsection{The gamma-ray spectrum}
\begin{figure*}
\includegraphics[scale=0.4,angle=0]{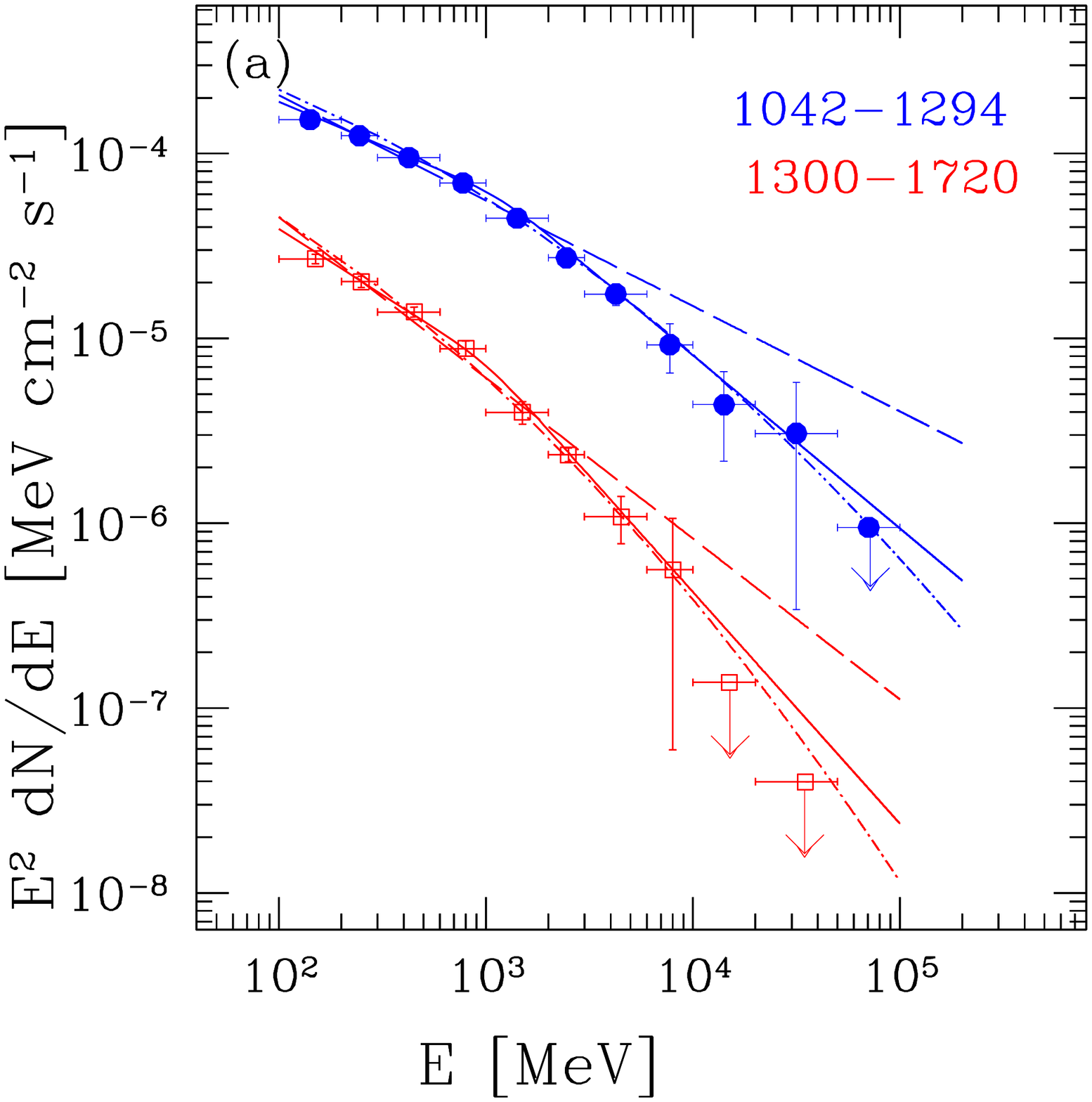}
\includegraphics[scale=0.4,angle=0]{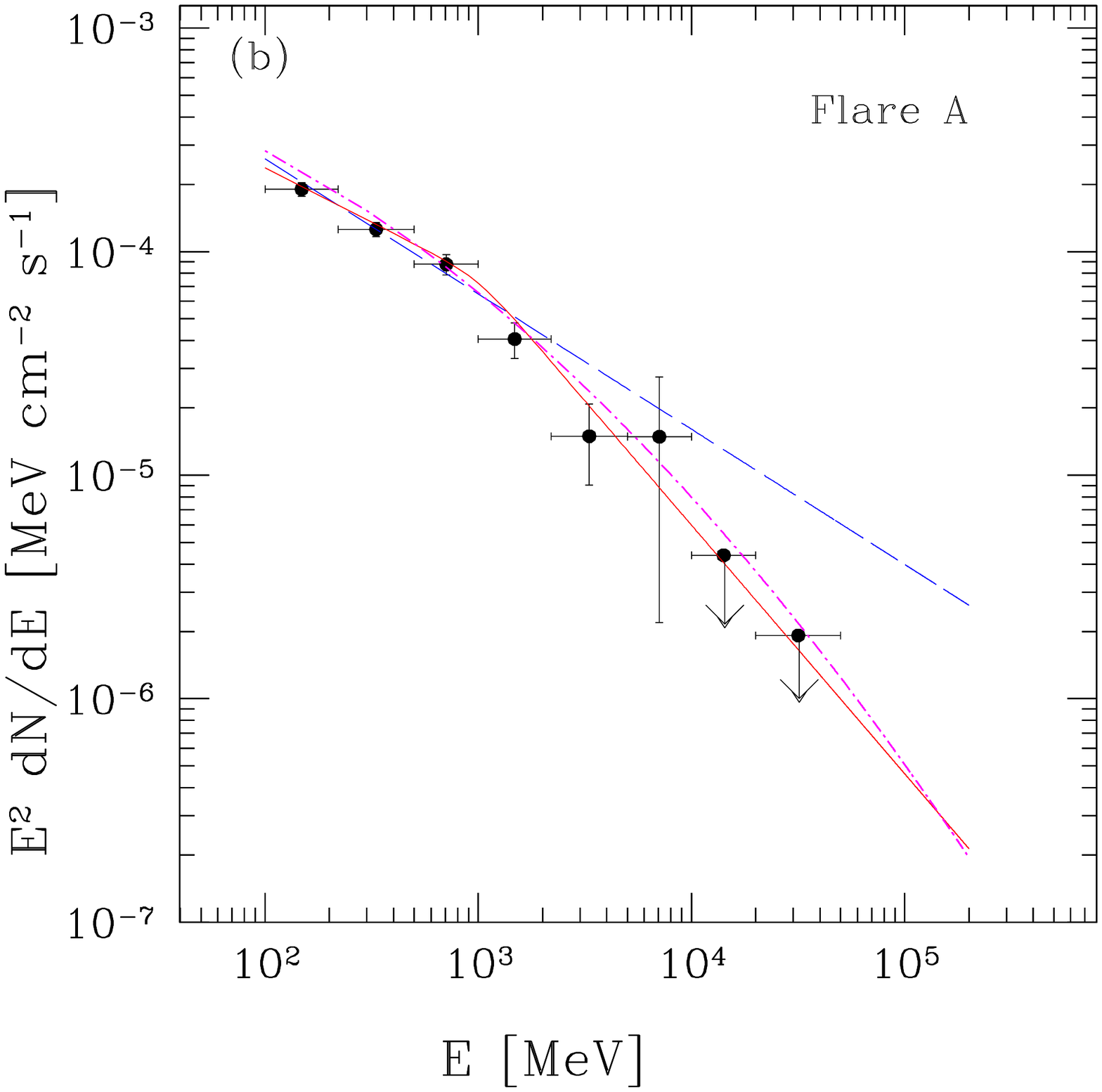}
\includegraphics[scale=0.4,angle=0]{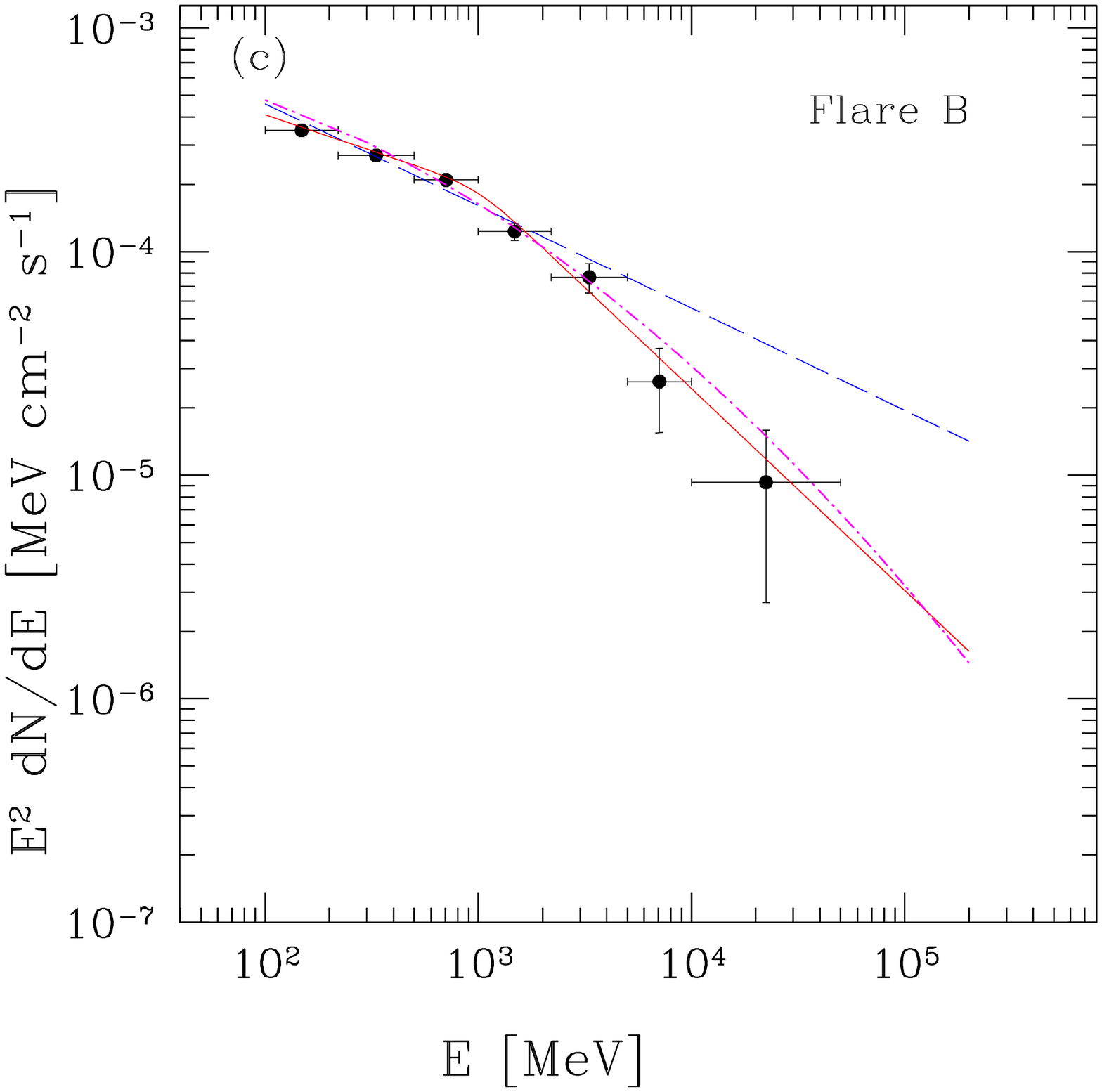}
\includegraphics[scale=0.4,angle=0]{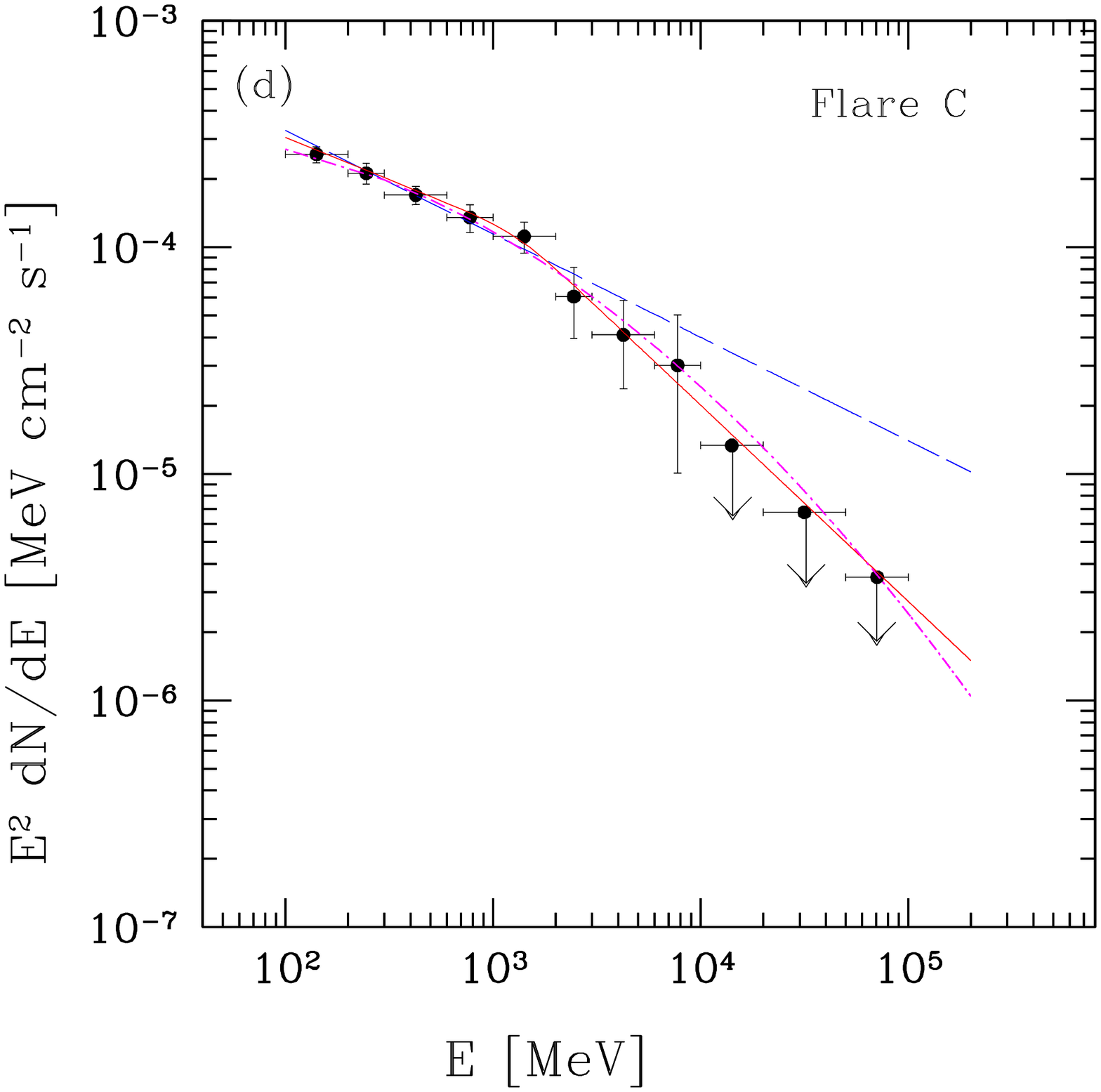}
\includegraphics[scale=0.4,angle=0]{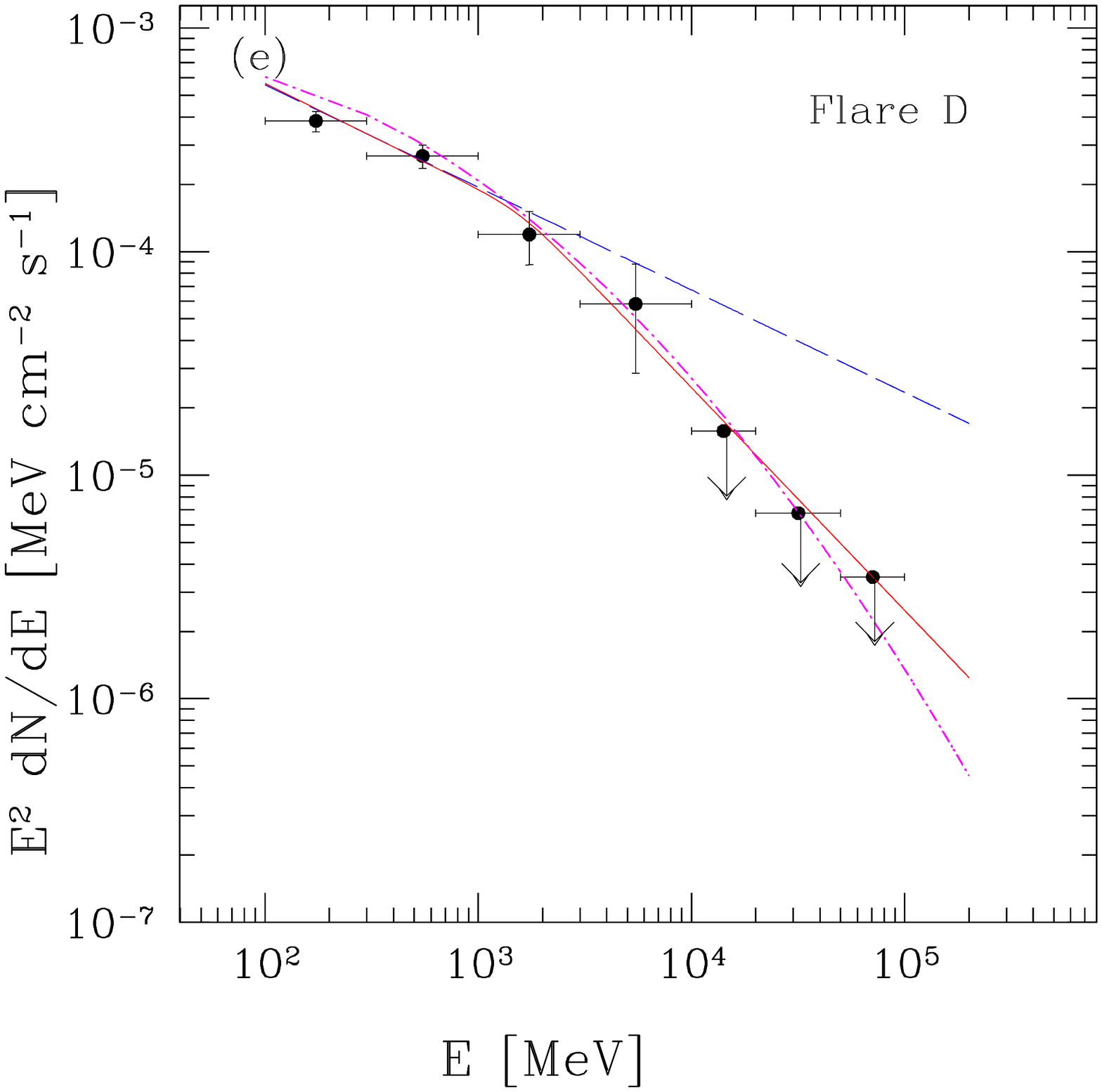}
\hspace{0.8in}
\includegraphics[scale=0.4,angle=0]{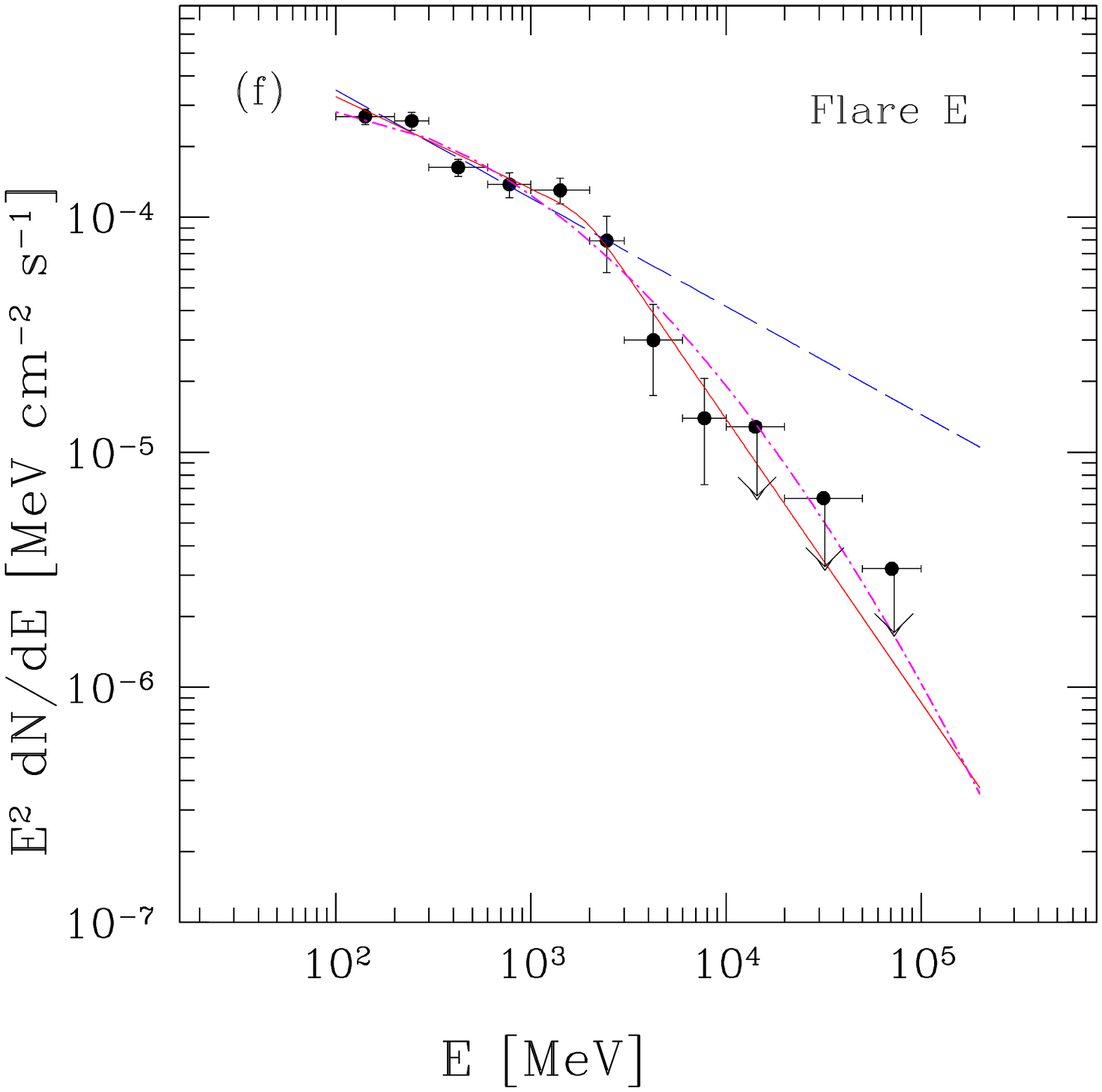}
\caption{Gamma-ray spectral energy distributions of 3C 273 during different activity
states along with the fitted PL (dashed curve), logP (dotted-dashed curve) and SBPL (solid curve) 
spectral models.   }
\label{spectra}
\end{figure*}

We construct the $\gamma$-ray spectrum using data for the entire flaring period (JD$^{\prime}$ = 1042 -- 1294) and for the 
quiescent state (JD$^{\prime}$ = 1300 -- 1720). 
Fig. \ref{spectra} (a) shows the $\gamma$-ray spectrum of the source over the high activity period and  the quiescent state. 
The observed spectrum shows a clear deviation from a 
simple power law. The spectral analysis using a broken power law model returns a break energy = 1.5$\pm$0.1 GeV with a 
likelihood test statistic relative to power-law, LRT = $-2 \Delta (ln~L)$ = 35.1 ( L is the value of the likelihood function), 
which corresponds to a significance of 
$\sim$5.58$\sigma$. The significance is estimated using Wilks' theorem i.e. the LRT  follows a chi-square 
distribution with $\Delta$n degrees of freedom ($\Delta$n is difference of number of degrees of freedom of the 
two spectral models). 
However, the spectrum does not exhibit a sharp break, rather the spectrum shows curvature near the break frequency.  
We therefore add a smoothing parameter $b$ to the broken power law 
model (see Section \ref{obs}) , which is fixed to 0.1\footnote{The results are not very sensitive to the choice of $b$ values.} while fitting.

We also investigate the spectral shape using the log-parabola (logP) model. For the high activity period, 
the LRT value of SBPL with respect to logP is 5.3, which corresponds to a significance of $\sim$2.3$\sigma$ 
with one degree of freedom. This implies that the SBPL model provides a marginally better fit compared to the logP model.  
The curves in Fig. \ref{spectra} (a) 
represent the fitted power laws i.e. PL, logP and SBPL 
in an energy range between 100 MeV to 300 GeV using the maximum likelihood method. 
The fitted model parameters over the energy range between 100 MeV to 300 GeV are summarized in Table \ref{tab2}. 
The difference of the logarithm of likelihood, $-2 \Delta L$ of SBPL and logP with respect to SPL is given in 
the second last column of Table \ref{tab2} along with the corresponding significance by which SBPL and logP are favored compared to the PL 
model. We therefore conclude that both smoothed broken power law and log-parabola models better 
describe the $\gamma$-ray spectral shape than the simple PL model with a break energy $E_{break}$ = 1.2$\pm$0.2 GeV. 
A similar indication of break at $\sim$1.5 GeV in the spectrum of 3C 273 integrated over the first 6 months of LAT 
observations has been also reported in \citet{abdo_spectra}.

The high activity period (JD$^{\prime}$ = 1042-1294) consists of five rapid flares (A, B, C, D, E). It is very likely that the physical 
conditions within the emission region change during different flares. We therefore investigate the
$\gamma$-ray spectra for the individual flares. 
Fig. \ref{spectra} (b)-(f) show the $\gamma$-ray spectra of the source for the individual flares. 
For each flare, we 
found that the spectrum shows a smooth rather than sharp break between  1 to 2 GeV.
As discussed in Section \ref{sec_lc}, the bright flares (A to E) are further composed of a number of sub-flares. 
Due to low photon statistics, the curvature in the spectra of the sub-flares can not be studied. 
Similarly, the absence of curvature was also noticed by \citet{abdo_3c273} during the two bright $\gamma$-ray sub-flares 
(sub-components of Flare B).

\begin{figure*}
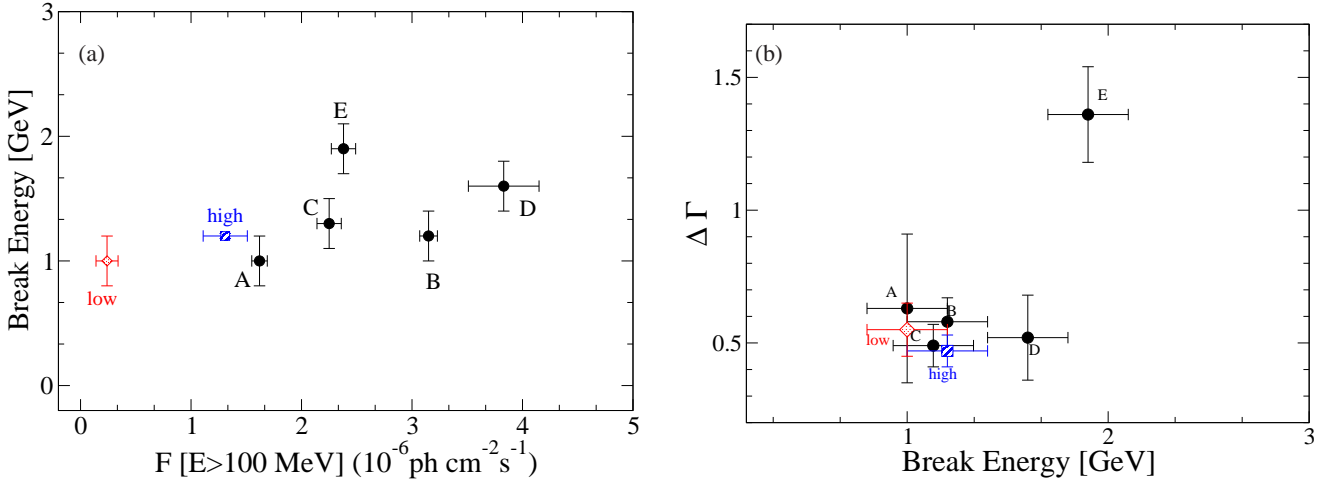

\vspace{0.2in}
\includegraphics[scale=0.35,angle=0]{Ebrk_flx.eps}
\put(-210,158){(a)}
\hspace{0.2in}
\includegraphics[scale=0.35,angle=0]{plt_G_Ebrk.eps}
\put(-210,158){(b)}
\caption{(a) Break Energy plotted as a function of flux for the different activity periods.  (b) Change of
the spectral slope $\Delta \Gamma$ as a function of break energy. 
In both plots, the squares represent the high activity period, the diamonds represent the low activity period, 
and the circles are the individual flares during the high activity period.
}
\label{Ebrk_flx}
\end{figure*}

The variation of break energy (calculated using SBPL) with the photon flux variations is displayed in 
Fig. \ref{Ebrk_flx} (a). 
As we see here, no strong variation in break energy with respect to the flux variation is found in the source.  
This is similar to other bright {\it Fermi} blazars \citep{abdo_3c454, ackermann2010, rani2013b}. 
The break energy $E_{break}$ varies within a factor of 2, while, the flux varies by a factor of $\approx$16. 
 Fig. \ref{Ebrk_flx} (b) 
displays the variation of $\Delta \Gamma$ = $\Gamma_2$ $-$ $\Gamma_1$ (see Section \ref{obs}) with the break energy ($E_{break}$). 
Again, the break energy remains nearly constant for a significant variation in $\Delta \Gamma$.

\begin{table*}
\scriptsize
\caption{ Parameters of fitted power laws }
\begin{tabular}{c c c c c c c c c r } \hline
Bin& JD$^{\prime}$  & $F_{100}$   & Model&$\Gamma$/$\alpha$/$\Gamma_1$ & $\beta$/$\Gamma_2$  & $E_{break}$ &$\Delta \Gamma$& $-2\Delta (ln~L)$ & Significance$^{a}$  \\
   &[JD-2454000] &(10$^{-6}$ ph cm$^{-2}$ s$^{-1}$)&     &    &      & (GeV)          &    ($\Gamma_2 - \Gamma_1$)           &  \\\hline
high     &1042-1294  &1.31$\pm$0.02 &PL &2.57$\pm$0.01  &              & ...       &     &    &                   \\
         &           &              &logP&2.74$\pm$0.03  &0.08$\pm$0.01 & ...   &             &44.7  & $ 6.7~ \sigma$       \\ 
         &           &              &SBPL&2.47$\pm$0.02  &2.94$\pm$0.06 &1.2$\pm$0.2&0.47$\pm$0.06&40.5 & $ 6.1 ~\sigma$ \\\hline 
low      &1300-1720  &0.24$\pm$0.10 &PL &2.87$\pm$0.05  &              & ...     &             &      &              \\
         &           &              &logP&2.86$\pm$0.04  &0.09$\pm$0.04 & ...         &             &9.9  &$ 3.3~\sigma$        \\
         &           &              &SBPL&2.70$\pm$0.01  &3.25$\pm$0.10 &1.0$\pm$0.2&0.55$\pm$0.10&11.9  & $ 3.3~\sigma$    \\\hline
Flare A  &1058-1078  &1.62$\pm$0.07 &PL &2.60$\pm$0.05  &              & ...         &             &      &              \\
         &           &              &logP&2.82$\pm$0.10  &0.11$\pm$0.04 & ...      &             &6.3  & $ 2.6 ~\sigma$       \\
         &           &              &SBPL&2.48$\pm$0.08  &3.11$\pm$0.28 &1.0$\pm$0.2&0.63$\pm$0.28& 9.8 & $ 2.9 ~\sigma$   \\\hline
Flare B  &1086-1114  &3.15$\pm$0.08 &PL &2.45$\pm$0.03  &              & ...         &             &      &              \\
         &           &              &logP&2.63$\pm$0.13  &0.12$\pm$0.06 & ...    &             &29.6  & $ $5.1 $\sigma$       \\
         &           &              &SBPL&2.32$\pm$0.02  &2.90$\pm$0.09 &1.2$\pm$0.2&0.58$\pm$0.09&24.9 &$ $4.7 $\sigma$ \\\hline
Flare C  &1140-1150  &2.25$\pm$0.11 &PL &2.45$\pm$0.04  &              & ...         &             &      &              \\
         &           &              &logP&2.56$\pm$0.09  &0.06$\pm$0.02 & ...    &             &7.3  & $2.9 ~\sigma$       \\
         &           &              &SBPL&2.37$\pm$0.03  &2.86$\pm$0.08 &1.3$\pm$0.2&0.49$\pm$0.08& 9.9 & $ 2.9~ \sigma$  \\\hline
Flare D  &1202-1204  &3.83$\pm$0.32 &PL &2.46$\pm$0.08  &              & ...         &             &      &              \\
         &           &              &logP&2.74$\pm$0.12  &0.25$\pm$0.06 & ...   &             &9.8  & 3.3 $\sigma$       \\
         &           &              &SBPL&2.47$\pm$0.10  &2.99$\pm$0.13 &1.6$\pm$0.2&0.52$\pm$0.16&6.3  & $2.2~ \sigma$       \\\hline
Flare E  &1260-1275  &2.38$\pm$0.11 &PL &2.46$\pm$0.02  &              &  ...        &             &      &              \\
         &           &              &logP&2.63$\pm$0.18  &0.09$\pm$0.07 & ...        &             &7.2  &2.7 $\sigma$        \\
         &           &              &SBPL&2.36$\pm$0.05  &3.72$\pm$0.18 &1.9$\pm$0.2 &1.36$\pm$0.18&12.1 & 3.1 $\sigma$    \\\hline
\end{tabular} \\
$\Delta (ln~L)$ is the difference of the -log(likelihood) value of logP, SBPL with respect to PL. \\
$a$ : We follow Wilks' theorem to estimate the significance i.e. twice the difference between the log(Likelihood) 
values for the two spectral models is formally distributed as $\chi^2$ with $\Delta$n degrees of freedom, where 
$\Delta$n is difference of number of degrees of freedom of the two spectral models.  
\label{tab2}
\end{table*}

\subsection{Gamma-ray Doppler factor}
The minimum Doppler factor $\delta_{\gamma}$ can be evaluated using the fact that the 
high-energy $\gamma$-ray photons can collide with the softer radiation to produce e$^{\pm}$
pairs. The cross-section of this process is maximized  
$\sim \sigma_{T}/5$ \citep[see][for details]{svensson1987}, where $\sigma_{T}$ is the Thompson 
scattering cross-section. This leads to  a lower limit on $\delta$ with the 
requirement that the optical depth $\tau_{\gamma \gamma} (\nu) < 1$ \citep{dondi1995, finke2008} :
\begin{equation}
\delta_{\gamma} > \left[ \frac{2^{a-1} (1+z)^{2-2a} \sigma_T d_L^2}{m_e
c^4 t_{var}} \epsilon f_{\epsilon^{-1}}^{syn} \right]^\frac{1}{6 - 2a}
\end{equation}
where $a$ is the power law index of synchrotron flux i.e. $f_{\epsilon}^{syn} \propto \epsilon^{a}$, and 
$m_e$ is the electron mass, $\epsilon = E/(m_e c^2)$ 
is the dimensionless energy of the $\gamma$-ray photon with energy E for which the optical depth of the emitting region
$\tau$ = 1.  The luminosity distance d$_L$ corresponding to z=0.158 \citep{strauss1992} is  
d$_L$\footnote{The results are not sensitive to d$_L$ = 0.784 Gpc for the recent Planck measurements 
of cosmology parameters \citep{plank2013}.} = 0.749 Gpc for a $\Lambda$CDM cosmology with $\Omega_m$ = 0.27, $\Omega_{\lambda}$ = 0.73, and
H$_0$ = 71 km s$^{-1}$ Mpc$^{-1}$ \citep{spergel2003}.
For the highest energy $\gamma$-ray (15.4 GeV)  photon observed in the source, we obtain 
$\epsilon$ = 15.4 GeV/(5.11 $\times 10^{-4}$ GeV) = $3 \times 10^4$ and $\epsilon^{-1}$ = $3.3 \times 10^{-5}$.  
Using $f_{\epsilon^{-1}}^{syn}$ = 4$\times 10^{-10}$ erg cm$^{-2}$ s$^{-1}$, observed by {\it Swift}/UVOT between 
January 10-13, 2010 \citep{giommi2012} and 
$t_{var}$ = 0.07 days (taken as the rising time of the most rapid flare D),  we obtain 
$\delta_{\gamma}$ $\geq$ 10.4.  
The estimated $\delta_{\gamma}$ value is  comparable to  $\delta_{VLBI}$ = 5--12 found by \citet{savolainen2006, jorstad2005} using 
kinematics of the parsec scale VLBI jet derived at radio wavelength.

\subsection{Size of the emission region}
One can also obtain an estimate to the size of the emission region ($R$) using the the calculated 
Doppler factor and variability time scale ($t_{var}$) i.e. $R \leq c ~t_{var} ~\delta_{\gamma}/(1 + z)$ 
\citep{rani2013a}. Using $t_{var}$ = 0.07 days, we obtain $R \leq 1.6\times10^{14}~ \delta_{\gamma} $ cm.
A lower limit of the estimated Doppler factor  $\delta_{\gamma}$ $\geq$ 10.4 gives $R \cong 1.6\times10^{15}$ cm. 
 
The angular size of the emission region is calculated using the following equation \citep{rani2013a} :  
\begin{equation}
\theta \leq {\rm 0.173\,\frac{t_{var}}{d_{L}}\,\delta_{\gamma} (1+z)\,\,  mas}  
\label{dim}
\end{equation}
where d$_L$ is is the luminosity distance in Gpc.  
Using $t_{var}$ = 0.07 days, d$_L$ = 0.749 Gpc, 
$\delta_{\gamma}$ $\geq$ 10.4, we obtain $\theta$ $\cong$ 0.52 $\mu$as which is much smaller than the size of the core region 
$\theta_{core}$ = (70$\pm$10) $\mu$as estimated using  VLBA observations \citep{savolainen2008}. 
It is worth pointing out that the $R$ and $\theta$ values found in this section could be smaller, since the variability time only gives upper
limits; or larger, since the Doppler factor used is a lower limit.

\section{Discussion}
\subsection{Rapid gamma-ray flares}
A prominent flaring activity was observed in the source between August 2009 and April 2010. The high activity 
period of the source was later followed by a quiescent state which is still continuing.  The fastest 
$\gamma$-ray flare observed in the source had a flux doubling timescale of 1.1 hr. 
Gamma-ray flares on similar time scales are also observed in many 
other bright {\it Fermi}-blazars \citep{abdo1502, foschini2011, saito2013, foschini2013, nalewajko2013, abdo_3c454}. 
Flux doubling times in the GeV band have  been reported in PKS1510$-$089 to be less than 3 hours \citep{saito2013} or  
as short as 20 minutes \citep{foschini2013}. Although these analyses did not  properly take into account the 
discontinuous LAT exposure pattern as  in our method, we conclude that the fastest flare observed in 3C 273  
is comparable to the most rapid flares seen in other {\it Fermi}-LAT  detected blazars.

In comparison to the asymmetric profile of very strong events of September 2009 (flare B), flares A, C, D and 
E are characterized by symmetric profiles. The asymmetric profile can be explored in terms of a fast injection of accelerated
particles and a slower radiative cooling and/or escape from the active region \citep{sikora2001}. 
The result of the superposition and blending of several episodes of short
duration could provide symmetric flare shapes \citep{valtaoja1999}. 
Although the authors only consider external-Comptonization in their model, it is worth pointing out that 
synchrotron self-Comptonization is also possible. 
\citet{nalewajko2013} argued that different Doppler boosting for different shells of the emission region could also be 
responsible for the asymmetry of $\gamma$-ray flares.

\subsection{Origin of spectral breaks}
Our analysis shows that 
the $\gamma$-ray spectrum of the source clearly deviates from a simple power law and is better described by a 
smooth break at $\sim$1.2 GeV. For the individual bright flares the break energy 
lies between 1--2 GeV. During the individual flares, the change in spectral slope above and below the 
break energy, $\Delta \Gamma$, varies between 0.47 to 1.36.

The observed spectral breaks in many bright {\it Fermi} blazars mostly lie within 2-10 GeVs \citep{poutanen2010, tanaka2011}.
Among the proposed scenarios,  
the absorption of $\gamma$-rays via photon-photon pair production on He II Lyman recombination continuum and 
lines  within the broad-line region (BLR) \citep[e.g.][and references therein]{poutanen2010, tanaka2011} can be 
responsible for the observed breaks.   
The $\gamma$-ray emitting region must be located within the BLR for this model to work.
Alternatively, the $\gamma$-ray spectral breaks could also be explained by a combination of two Compton-scattered 
components, for example, Compton scattering of the disk and BLR radiation as proposed by \citet{finke2010}. 
Dermer et al. (2013) proposes a 3-parameter log-parabola model for the electron energy distribution, which 
naturally produces a GeV spectral cutoff by scattering of Ly-$\alpha$ and BLR radiation.
Another explanation refers to an intrinsic origin of the spectral breaks. A change in spectral index below and 
above the break of order $\sim$0.5 can be interpreted as a typical ``cooling break" associated with radiative 
losses \citep{abdo2009}. The observed softening may instead be due to an intrinsic decline or break in the particle 
distribution also.

For 3C 273, the change in spectral slope ($\Delta \Gamma$) above and below the break energy varies 
between 0.47 to 1.36. The estimated $\Delta \Gamma$ values for flare E with $\Delta \Gamma$$\sim$1.36, do not favor 
the standard radiative cooling models that predict a spectral break of 0.5 units. Although for flare A to D, 
$\Delta \Gamma$ is close to 0.5, it is difficult to reconcile the constancy of the break energy with respect 
to the photon flux variations within the ``cooling break" scenario. 
This does not argue in favor of the intrinsic origin of the observed spectral breaks in 3C 273 
associated with the radiative cooling.

If the spectral breaks are due to $\gamma \gamma$ absorption, then one can estimate the frequency of 
the most efficiently interacting target photons with $\gamma$-rays. 
For head-on collisions, we have $\nu \approx 6 \times 10^{14} (E/100 ~GeV)^{-1}$ Hz. 
A spectral break at $E_{break}$ = 1--2 GeV in the observer frame 
corresponds to 0.86--1.73 GeV ($E_{break}$/(1+z)) in the source frame.  
Therefore, for the spectral break within 0.86--1.73 GeV, the target photons fall within (3.4--6.9)$\times 10^{16}$ Hz i.e. are extreme 
ultra-violet photons. This implies that the target photon energy is higher than  the He II and H I recombination continuum. 
Moreover, spectral breaks close to 1 GeV are not very likely to be observed within this 
scenario because of low opacity at these energies \citep{poutanen2010}.

The optical-UV spectrum of the FSRQ 3C 273 shows a prominent excess of emission, which is mainly interpreted 
as  a  contribution of the accretion disk emission 
\citep[see][and references therein]{ulrich1981, soldi2008}  or the presence of a hot corona particularly in the UV-Xray 
regime \citep{haardt1994}. 
The excess optical-UV  emission can be approximated by a black body with a temperature of $\sim$21,000--26,000 K with 
a characteristic dimension of 10$^{16}$ cm \citep{ulrich1981, pian1999}. 
Consequently, Compton processes of a thermal plasma in the disk or in the corona  can contribute efficiently  
to the high-energy radiation. The optical-UV spectrum of 3C 273 also shows strong emission lines e.g. Ly-$\alpha$, 
CIV, OVI, CIII, NIII, and SVI \citep[e.g.][and references therein]{appenzeller1998, paltani2003}. 
Therefore it might not be surprising  to find a contribution of inverse-Compton scattered radiation from both 
the disk and the BLR.

\subsection{Location of the gamma-ray emission region}
Assuming a conical geometry, the observed variability timescale of the source can also be used to put an upper limit on the 
distance of the emission region (r$_{\gamma}$) from the central engine \citep{tavecchio2010} using 
r$_{\gamma}$ $<$ (t$_{var}$\,c\,$\delta$)/($\theta_{jet}$\,(1+z)); where $\theta_{jet}$ is the opening angle 
of the jet. Moreover, as we are using a lower limit of Doppler factor $\delta_{\gamma}$, the 
estimated value of $r_{\gamma}$ is supposed to be close to the actual value.    
Using the rising time of flare D, t$_{var}$ = 0.07 day and $\delta_{\gamma}$ =10 and 
$\theta_{jet}$ $\sim$1.4$^{\circ}$ \citep{jorstad2005}, we obtain r$_{\gamma}$$\sim$0.03 pc. 
We however stress that this is a very rough assumption as the jet may not be conical \citep{hada2012, krichbaum2006}.
Moreover, the $\gamma$-ray emitting region of the jet could be much closer to the black hole (with a different Doppler factor) 
compared to the portion of the jet resolved by the VLBI \citep{rani2013a}.

The energy dependence of the cooling timescale can also be used as an alternative approach to constrain the 
location of the $\gamma$-ray emission region \citep{dotson2012}. 
For 3C 273, we found that the cooling timescales of the bright $\gamma$-ray flares are similar for $E < 1$ GeV 
and $E > 1$ GeV light curves with a possible minimum temporal resolution of 6 hr. This implies that the allowed difference 
for $F_{E<1GeV}$ and $F_{E>1GeV}$ decay timescale  is shorter than 6 hr. An upper 
limit for the distance of the $\gamma$-ray emission  region from the central engine is given by 
$r_{\gamma}$ $<$ 2.3$\times$10$^{19}$ $\Gamma_{bulk}$ [$\Delta t_{max,hr}~ L_{MT,45}/(1+z)^{1/2}$]$^{1/2}$ cm, where $L_{MT,45}$ 
is the molecular torus luminosity in units of 10$^{45}$ erg s$^{-1}$  and $\Gamma_{bulk}$ is the bulk 
Lorentz factor \citep{dotson2012}. The molecular torus luminosity can be considered as a 0.1-0.5 fraction of the
accretion disk luminosity (L$_{AD}$). For 3C 273, we have L$_{AD}$$\sim$1.6$\times$10$^{46}$ erg s$^{-1}$ 
\citep{kriss1999, soldi2008, giommi2012}. 
Using L$_{MT,45}$$\sim$1.6, $\Gamma_{bulk,10}$ = 10.6$\pm$2.8 \citep{jorstad2005}  
and $\Delta t_{max}<$ 6 hr, we obtain $r_{\gamma}<$ 1.6 pc.

\citet{finke2010} argue that a region where the Compton-scattered disk and BLR
emissions are approximately equal can explain the {\it Fermi}-LAT emission in 3C 454.3. Their 
analysis implies that the location of $\gamma$-ray emission region ($r_{\gamma}$) is $R_i < r_{\gamma} << \Gamma^{4}_{bulk} r_g$, 
where $R_i$ is the inner radius of BLR, $\Gamma_{bulk}$ is the bulk Lorentz factor and 
$r_g$ is Schwarzschild radius of the central black hole.  
They also propose a robust solution where 
the external field energy density from disk and BLR radiation decays as $\sim r^{-3}$. 
In such a scenario, a comparison of BLR and disk radiation gives $\tau_{BLR} = (R_i/r_g)^{-1}$ ($\tau_{BLR}$ is the Thomson
depth of the BLR) for the emission regions formed in the BLR. 
For 3C 273, $M_9 = M_{BH}/(10^9 M_{\odot}$) $\approx$ 0.9-2.4 \citep{paltani2005,peterson2004} where
$M_{BH}$ is the black hole mass. 
Using $\tau_{BLR}$ = 0.01 to 0.1 \citep{bottcher1998}, and $\Gamma_{bulk}$ = 10.6$\pm$2.8 \citep{jorstad2005},  we obtain 
0.005 pc $<$ $r_{\gamma} <$ 1.4 pc. 
The achromatic cooling of flares constrains $r_{\gamma}$ within 1.6 pc.   
Therefore the allowed range of $r_{\gamma}$ using the two methods is 0.005 pc $<$ $r_{\gamma} <$ 1.6 pc.

Our analysis suggests that the location of the $\gamma$-ray flaring activity observed in 3C 273 lies within 
1.6 pc distance from the central engine. The achromatic cooling of the $\gamma$-ray flares below and above 
1 GeV is consistent with the $\gamma$-ray emission region being located within the BLR. Also, the observed spectral 
breaks (at $\sim$1-2 GeV) can be well described by $\gamma$-ray absorption 
within the BLR. Moreover, as a result of $\gamma \gamma$ pair production, the BLR of FSRQs is opaque to $\gamma$-rays 
above $\sim$20 GeV, while the MT region is not \citep{donea2003}. As we saw in Section 3.1, the highest energy 
$\gamma$-ray photon observed for 3C 273 has energy 15.4 GeV, and provided the fact that the source has not been detected 
at TeV energies so far, therefore our results are consistent  with the $\gamma$-ray emission site being located within 
the BLR.

\subsection{The orphan flare E}
We found that the $\Delta \Gamma$ for flare E ($\sim$1.5) is significantly different from 
that for flares A to D ($\sim$0.5). Note that flare E is the last flare observed during the high activity period. 
Also, the break energy of flare E (E$_{break}$$\sim$2 GeV) is higher than for the rest of the flares. Moreover, the 
doubling time scales of flare E are comparatively longer than those for flares A to D. We also notice that in the 
monthly averaged light curves (Fig. \ref{lc_flx}), flares A to D are the sub-components of a single flaring event, 
while flare E seems to be an independent flaring event.

The impact of the geometry of the BLR on the expected absorption, through the $\gamma \gamma$ process was
recently discussed by \citet{tavecchio2012}. They argued that if the BLR has a full covering factor, the break energy 
does not change as long as the emission occurs within the BLR, but $\Delta \Gamma$ decreases as the emission region moves away 
from the central engine. A correlated $\Delta \Gamma$ - $E_{break}$ variation is expected for the partially covered BLR. 
For flare A to D, the nearly constant values of $\Delta \Gamma$ and $E_{break}$ suggest  
an approximately fixed location of the emission region from the black hole. 
The higher $\Delta \Gamma$ value for flare E infers a closer location to the central engine compared to flare A to D. 
However, the longer variability timescale for flare E could argue for an emitting region larger and farther out 
(if the jet is not stratified), contrary to the result from the larger $\Delta \Gamma$.

With the recent jet kinematics study of 3C 273 at 
7 mm wavelength, \citet{jorstad2012} reported an ejection of four components during the high $\gamma$-ray  activity 
period between mid-2009 to mid-2010. The authors identified two components during the high $\gamma$-ray activity 
period. The ejection of the fastest knot was associated with the most prominent $\gamma$-ray peak (flare B), while the brightest component
had maximum flux coinciding with flare E. This suggests that the flares A-D and E are independent flaring events. 
Multiple shock scenarios and/or shock-shock interactions are among the feasible 
mechanisms for these flaring events \citep{sokolov2004,fromm2012}. 
It could also be possible that flare E is located at similar distances from the central engine than flare A-D,
but is in a different layer or sheath in the jet. Such a scenario is expected within the multi-zone
emission model of \citet{marscher2013}, which describes a stratified jet.

\section{Conclusions}
The continuous monitoring in the high-energy $\gamma$-ray band by the {\it Fermi}-LAT allows us to investigate the
$\gamma$-ray flux and spectral variability of the FSRQ 3C 273. The source displays prominent flaring activity during 
August 2009 to April 2010. Five rapid flares are observed in the source during the interval. 
The rapid flares recur roughly at 50 day intervals. Each flare is further composed of sub-flaring components with 
characteristic rise and decay timescales as fast as few hours. The fastest $\gamma$-ray flare observed in the source 
has a characteristic flux doubling time scale of 1.1 hr and is compatible with the fast $\gamma$-ray flares observed in other
{\it Fermi} blazars. The 3 hr peak flux of the flare above 100 MeV is (12.3$\pm$2) $\times$10$^{-6}$ ph cm$^{-2}$ s$^{-1}$, 
corresponding to an apparent isotropic $\gamma$-ray luminosity of 2.6$\times$10$^{46}$ erg s$^{-1}$. 
The source exhibits a strong evolution of spectral index ($\Gamma$) during the different activity states.

The highest energy $\gamma$-ray photon (15.4 GeV) observed for the source 
arrives during this high-activity period. 
The minimum $\gamma$-ray Doppler factor, $\delta_{\gamma}$ derived using $\gamma \gamma$-opacity constraints 
for the highest energy $\gamma$-ray photon is 10, which is comparable to $\delta_{VLBI} \sim$5--11 obtained using 
jet kinematics \citep{savolainen2006, jorstad2005}.
Causality arguments constrain the size of the emission region to 1.6$\times10^{15}$ cm.

The $\gamma$-ray spectra measured over this period show clear deviation from a simple power law with a 
break in the 1-2 GeV energy range. During different flaring epochs, no strong evolution 
of E$_{break}$ is found. We have discussed possible explanations for the origin of $\gamma$-ray spectral breaks. 
The different possible scenarios allow us to constrain the location of $\gamma$-ray emission region within 
the BLR ($<$1.6 pc). We argue that the observed $\gamma$-ray emission in 3C 273 is unlikely to have a contribution 
of the molecular torus photons in external-Comptonization. 
We found that the spectral parameters (E$_{break}$ and $\Delta \Gamma$) of flare E are not comparable to the 
spectral parameters of flares A to D. A comparison of the $\gamma$-ray flares with the jet kinematics \citep{jorstad2012} suggests 
that flare E has a different physical origin. A detailed discussion of physical mechanisms, along with results of the study 
focused on broadband flux variability and spectral modeling of the source will shed more light on this, and will be 
given in a separate paper (Rani et al. 2013c in preparation).

\section*{Acknowledgments}
The {\it Fermi}-LAT Collaboration acknowledges generous ongoing support
from a number of agencies and institutes that have supported both the
development and the operation of the LAT as well as scientific data analysis.
These include the National Aeronautics and Space Administration and the
Department of Energy in the United States, the Commissariat{\'a} l'Energie Atomique
and the Centre National de la Recherche Scientifique / Institut National de Physique
Nucl\'eaire et de Physique des Particules in France, the Agenzia Spaziale Italiana
and the Istituto Nazionale di Fisica Nucleare in Italy, the Ministry of Education,
Culture, Sports, Science and Technology (MEXT), High Energy Accelerator Research
Organization (KEK) and Japan Aerospace Exploration Agency (JAXA) in Japan, and
the K.~A.~Wallenberg Foundation, the Swedish Research Council and the
Swedish National Space Board in Sweden.
Additional support for science analysis during the operations phase is gratefully
acknowledged from the Istituto Nazionale di Astrofisica in Italy and the Centre 
National d'\'Etudes Spatiales in France. 

We would like to thank Stefano Ciprini, the internal referee for the {\it Fermi}-LAT Collaboration, 
Stefanie Komossa, Seth Digel, Marco Ajello and Justin Finke for their useful suggestions and comments. 
We thank the referee for constructive comments that have helped us to improve the paper.
BR gratefully acknowledges the travel support the COSPAR Capacity-Building Workshop fellowship program.

\label{lastpage}

\end{document}